\documentstyle[11pt]{article}
\newcounter{multieqs}

\begin{document}
\hfill {\Large DTP-MSU 92-01, November 1992}
\vspace{5cm}
\begin{center}
{\LARGE STRINGY SPHALERONS AND NON-ABELIAN BLACK HOLES}
\end{center}
\begin{center}
        {\bf E.E. Donets } and {\bf D.V.Gal'tsov}\\[0.2cm]
\begin{em}
         Department of Theoretical Physics, Physics Faculty,\\
         Moscow State University, 119899 Moscow, Russia\\
\end{em}
\end{center}

\vspace{2.5cm}

\begin{center}
{\bf Abstract}
\end{center}
Static spherically symmetric asymptotically flat particle-like
and black hole solutions are constructed within the SU(2) sector
of the 4-dimensional heterotic string effective action.
They separate topologically distinct Yang-Mills vacua and
are qualitatively similar to the Einstein-Yang-Mills
sphalerons and non-abelian black holes discussed recently. New
solutions possess quantized values of the dilaton charge.

\vfill \eject

Ten-dimensional heterotic string theory seems to provide a
reasonable background for the discussion of the gravitational
dynamics on a microscopic scale. Recently it was shown that
an effective action for graviton, dilaton, axion and Yang-Mills
fields resulting from vanyshing of the relevant beta-functions
possess interesting (4-dimensional) black hole solutions \cite{r1}
that can shed new light onto the problem of a final stage of the
Hawking evaporation \cite{r2} (for a review see e.g. \cite{r3}).
New features of black holes in this model are related to the presence
of a dilaton. As far as the Yang-Mills content is concerned, they
belong to the U(1)-sector, i.e. essentially {\em abelian}.

 From the other hand, new {\em non-abelian} particle-like \cite{r4}
and black hole \cite{r5} solutions were discovered in the pure
Einstein-Yang-Mills (EYM) coupled system with the SU(2) and
higher rank groups \cite{r6}. They are generically unstable
and satisfy (for the SU(2) case) a non-abelian baldness theorem
\cite{r7} stating that within the class of static
asymptotically flat solutions only the embedded abelian ones
can possess conserved (electric or magnetic) Coulomb charges.
These features were further related to the {\em sphaleron} nature
of the particle-like EYM solutions \cite {r8}, \cite{r9}.
The odd-n (number of zeros of the Yang-Mills potential) members
of the family were shown to lie on the top of the potential
barrier separating neighbouring topologically distinct Yang-Mills
vacua \cite{r8}. General Morse-type argument in favor of this
interpretation as well as some predictions concerning the number
of unstable modes were given in \cite{r9}.

A step forward to incorporate the EYM sphalerons into the
realistic heterotic string context was made recently by
Lavrelashvily and Maison \cite{r10} (see also \cite{r11}) who
found numerically particle-like solutions of the Yang-Mills-dilaton
coupled system in a flat space-time. The decoupling of gravity,
however, corresponds to the unphysical limit of a characteristic
parameter of the underlying string theory. Here we investigate
the problem using the full bosonic part of the heterotic
string effective action. We formulate (without proof) the
corresponding non-abelian baldness conjecture and show
(numerically) the existence of both regular and black hole
static asymptotically flat solutions in the SU(2) sector.
They (necessarily) have dilaton charges and thus possess
a Coulomb-type hair. This charge, however, is not a conserved
one, but rather is a depended quantity which is determined by the
distribution of the Yang-Mills field of the configuration.
Nevertheless, it is a new physical parameter describing
the interaction between two sphalerons (black holes)
due to the dilaton field. It is worth to be noted
that Gibbons abelian dilatonic black holes \cite{r1} share
the same property, that case a dilaton charge being
expressible in terms of the electric and magnetic charges.
We derive here a sum rule for the dilatonic charge which
reduces to the Gibbons identity in the abelian case and
which is also applicable to non-abelian configurations.

Other properties of the stringy non-abelian solutions
are very similar to those of the EYM counterparts: they form
a discrete sequence labeled by the number of nodes of the
Yang-Mills potential and exist for an arbitrary value of the
radius of the event horizon (regular solutions being a
limiting case). They have vanishing Chern-Simons 3-forms both
in the Lorentz and the Yang-Mills sectors and consequently a
Kalb-Ramond field is not excited (although a purely
topological axion charge is allowed).

We start with the following bosonic part of the 4-dimensional
heterotic string effective action in the Einstein frame
\begin{equation}
S = \frac{1}{16\pi} \, \int \, \sqrt{-g}\, [\, m^2_{Pl}\,(- R
-1/3 \exp(-4\Phi)\,H_{\mu\nu\lambda}H^{\mu\nu\lambda} \,+\, 2
\partial_{\mu} \Phi \partial^{\mu} \Phi) -
 \exp (-2 \Phi) \, F_{a\mu\nu} \, F_a^{\mu\nu}]\, d^4x \, ,
\end{equation} 
\noindent
where $\Phi$ is the dilaton, $H_{\mu \nu \lambda}$ is the Kalb-Ramond
fieid coupled to the Lorentz and the  Yang-Mills Chern-Simons 3-forms
$H=dB+\omega_{3L}-\omega_{3YM}$, $F_{a \mu \nu}$ is the Yang-Mills
curvature corresponding to some gauge group containing  SU(2) as a
subgroup (in what follows we shall deal with the SU(2) component
only). In terms of the Peccei-Quinn axion $a$
\begin{equation}
H^{\mu \nu \lambda}\,=\,1/2\, E^{\mu \nu \lambda \tau} \partial_{\tau} a
\, exp(4\Phi)
\end{equation} 
\noindent
the action (1) reads
\begin{eqnarray}
S  =  \frac{1}{16\pi}\, \int \, \sqrt{-g}\, [\,m^2_{Pl}\,(-R+2
\partial_{\mu} \Phi \partial^{\mu} \Phi + 1/2\, \partial_{\mu} a\,
\partial^{\mu} a \, exp(4\Phi)) - \nonumber\\
 -  exp (-2\Phi)\,F_{a\mu\nu}F_a^{\mu\nu}\, + \, a \,F_{a\mu\nu}
\tilde{F}_a^{\mu\nu}\,]\; ,
\end{eqnarray} 
\noindent
where $\tilde{F}$ is the dual field strength. Using the identity
$F_{a\mu\nu}\tilde{F}_a^{\mu\nu}=\nabla_{\mu}\,K^{\mu}$, where $K^{\mu}$
is the topological current dual to the Yang-Mills Chern-Simons 3-form,
and eliminating the {\em total divergence} one can cast the last term
in the Eq.(2) into the form $\nabla_\mu a\, K^\mu$ . After this an
axion field will enter into the action only under the gradient and
hence can be integrated out.

We are interested in the static spherically symmetric configurations
which can conveniently be described by the line element
\begin{equation}
ds^2=\frac{\Delta \sigma^2}{r^2} dt^2 - \frac{r^2}{\Delta}dr^2
 -r^2 (d\theta^2 + \sin^2 \theta d\phi^2)\,.
\end{equation} 
\noindent
The corresponding Yang-Mills connection after fixing the gauge can be
parameterized in terms of two real-valued functions $W(r)$ (electric
part) and $f(r)$ (magnetic part) as follows
\begin{equation}
gA_{a\mu} T^a dx^\mu = W L_r dt + (f-1)(L_\phi d\theta - L_\theta
\sin \theta d\phi)\, ,
\end{equation} 
\noindent
where $L_r =T^a n^a$,  $L_\theta=\partial_\theta L_r$,
$L_\phi= (\sin \theta)^{-1}\partial_\phi L_r\;,$
$n^a = (\sin \theta \cos \phi, \sin \theta \sin \phi, \cos \theta)$
is a unit vector, $T^a$ are normalized hermitean generators of SU(2),
and $g$ is the coupling constant.

After dimensional reduction and elimination of a total
divergence the action (3) will read
\begin{eqnarray}
S& =& \frac{1}{2} \int dt dr \{m^2_Pl [\sigma'(\Delta/r - r)
- (\Phi'^2 + a'^2 \exp(4\Phi))\Delta \sigma] +\nonumber\\
&+& g^{-2}[(\frac{W'^2 r^2}{\sigma} + \frac{2f^2 r^2 W^2}
{\Delta \sigma} - \frac{2f'^2 \Delta \sigma}{r^2}
- \frac{\sigma(1-f^2)^2}{r^2}) \exp (-2\Phi) - 2a'W(f^2-1)]\}\,,
\end{eqnarray} 
\noindent
where primes denote the derivatives with respect to $r$.
In the abelian sector $f\equiv 0$ it is equivalent to that considered
in \cite{r1} and generates the embedded abelian solutions similar
to Gibbons electrically and magnetically charged black holes with
long-range dilaton hair. Within the abelian sector
there exist generalized duality transformations which allow to
eliminate an axion field $a$, or, inversely, to generate solutions
with axion hair from Maxwell-dilaton ones.

The Yang-Mills part of the action (6) is close to that of the pure
EYM system \cite{r4}, \cite{r5}, \cite{r7} and can be analysed along
the same lines. We formulate here without proof the main result as
the {\em generalized non-abelian baldness} theorem:
\begin{em}
among all asymptotically flat configurations (both regular or
black hole) only embedded abelian ones can have electric or magnetic
charges.
\end{em}
Morover, one can show that solutions with $W\neq0$ necessarily have
an electric charge. So, to investigate the essentially
non-abelian solutions, we are led to consider only $W\equiv0$ case.
Then it can be easily shown that the axion hair have to vanish apart from
the purely topological hair of the type \cite{r12}. Indeed, a variation
of (6) with respect to $a$ leads to the following equation
\begin{equation}
a' = \frac{2 R^2_g W (1-f^2) \exp(-4\Phi)}{\Delta \sigma} \, + \, a_1\,,
\end{equation} 
\noindent
where $R_g=1/(m_{Pl}g)$ is the characteristic length, and $a_1= const$
results after the first integration. For $W=0$ the solution reads
\begin{equation}
a = a_0 + a_1 r\,,
\end{equation} 
\noindent
where $a_0 = const$. From the asymptotic flatness $a_1 =0$, while
$a_0$ can still be non-zero and be interpreted as a topological
axion charge \cite{r12}. Obviously such an axion hair is fully
non-dynamical and has no influence on the remaining fields configuration.

In the pure magnetic sector the variation of the action (6) gives
the following set of equations for $f, \Phi, \sigma$ and $\Delta$:
\begin{eqnarray}
(\frac{f' \Delta \sigma \exp(-2 \Phi)}{r^2})' =
\frac{\sigma f(f^2-1)\exp(-2\Phi)}{r^2} \; ,\\
(\Phi' \Delta \sigma)' = - \frac{R_g^2 \sigma F \exp(-2 \Phi)}
{r^2}\;,\\
(\ln \sigma)' = \frac{2R_g^2 f'^2 \exp(-2\Phi)}{r} + r \Phi'^2\;,\\
- (\frac{\Delta}{r})' + 1 = \frac{R_g^2 F \exp(-2\Phi)}{r^2} +
\Delta \Phi'^2 \;,
\end{eqnarray} 
\noindent
where
\begin{equation}
F=2\Delta f'^2 + (1-f^2)^2 \;.
\end{equation} 

The Yang-Mills equation (9) is scale-invariant and remains the same
under the constant shift of the dilaton $\Phi \rightarrow \Phi + c$.
All other equations recover their initial form after the accompanying
rescaling $R_g \rightarrow R_g e^c $. It is convenient to
fix the scale by imposing on the dilaton field an asymptotic condition
$\Phi (\infty) = 0$. Then it can be shown that the leading term for
the dilaton at infinity will be a Coulomb one
\begin{equation}
\Phi = \frac{D}{r} + O(\frac{1}{r})
\end{equation} 
\noindent
provided the Yang-Mills field is not in the vacuum state $|f|\equiv 1$.
It can be easily seen from the Eq.(11) that the dilaton charge $D$
dominates the asymptotic behavior of the metric function $\sigma$
\begin{equation}
\sigma = 1 - \frac{D^2}{2r^2} +O(\frac{1}{r^4})\;.
\end{equation} 

The asymptotic of the second metric function $\Delta$ has the form
\begin{equation}
\Delta = r^2 - 2Mr + D^2 + O(\frac{1}{r})
\end{equation} 
\noindent
where $M$ is the Schwarzschild mass. It should be noted that the
contributions of the dilaton charge squared is canceled in the
asymptotic behavior of the metric component
\begin{equation}
g_{00} = \frac{\Delta \sigma^2}{r^2} = 1 -\frac{2M}{r} +
O(\frac{1}{r^3})
\end{equation} 
\noindent
and hence the dilaton charge does not produce the Reissner-Nordstrom
type contribution to the asymptotic form of a metric.

An asymptotic behavior of the Yang-Mills function $f$ compatible
with the asymptotic flatness is either $f(\infty)=0$ which
corresponds to the magnetically charged configuration (abelian),
or $f=\pm 1$ (non-charged sphaleronic configuration)
\cite{r7}, \cite{r8}. Here we concentrate on the second case.

To specify the behavior of variables on the other side of the
$r$-semiaxis it is convenient to introduce the parameter $r_0$
which is zero for the regular sphaleron solution and is equal
to the horizon radius $r_H$ in the black hole case. In both
cases $\Delta(r_0)=0$. The other functions have at this point
finite and non-zero values. Then, integrating the dilaton
equation (10) from $r_0$ to infinity we obtain the following
sum rule for the dilaton charge
\begin{equation}
D= R_g^2 \int_{r_0}^{\infty} \frac{\sigma  F \exp(-2\Phi)}{r^2} dr
\end{equation} 
Since the function $F$ given by the Eq.(13) is positive definite
(as well as $\sigma$), the dilaton charge is non-zero for both
the abelian $f\equiv 0$, $\sigma = (1+D^2/r^2)^{-1/2}$ and the
non-abelian configurations. In the first case the Eq.(18) reproduces
the result of Gibbons \cite{r1} $D\sim (magnetic \; charge)^2$.

An analogous sum rule can be obtained for the Schwarzschild
mass by integrating the Eq.(12)
\begin{equation}
M=\frac{1}{2}\int_{r_0}^{\infty}(\Delta \Phi'^2 + \frac{F
\exp(-2\Phi)}{r^2})dr + M_0
\end{equation} 
\noindent
where $M_0=0$ in the regular case and $M_0=M_H$ -{\em ''bare``}
black hole mass in the black hole case.

An equation for the metric function $\sigma$ (11) can be
integrated as follows
\begin{equation}
\sigma=\exp[ - \int_r^{\infty} (\frac{2R_g^2 f'^2 \exp(-2\Phi)}
{r}+r \Phi'^2)dr]
\end{equation} 
and then substituted into the other equations (9), (10), (12).
Transforming to a new variable $\rho=r^2$ one obtains finally
the following set of coupled equations for three quantities
$f$, $\Phi$ and $\Delta$
\begin{eqnarray}
\Delta  (f_{\rho \rho} -2f_\rho \Phi_\rho) + \frac{1}{2}G f_\rho +
\frac{f(1-f^2)}{4\rho} = 0 \;,\\
\Delta  (\Phi_{\rho \rho} +\frac{1}{\rho} \Phi_\rho) +
\frac{1}{2} G \Phi_\rho + \frac{R_g^2F\exp(-2\Phi)}{4\rho^2}=0 \;,\\
\Delta_\rho  +\Delta(2\rho\Phi_\rho^2 -\frac{1}{2\rho})+\frac{1}{2}
(\frac{R_g^2F\exp(-2\Phi)}{\rho}-1)=0 \;,
\end{eqnarray} 
\noindent
where
\begin{equation}
G = 1 - \frac{R_g^2(1-f^2)^2\exp(-2\Phi)}{\rho}\,.
\end{equation} 

First we turn to the discussion of the regular solutions. A series
expansion of the solution of the system (21)-(23) in the vicinity
of the origin can be written as follows
\begin{eqnarray}
f&=& -1 + b x + O(x^2)\,,\\
\Phi& =& \Phi_0 - 2  b^2\,x\, \exp(-2\Phi_0) + O(x^2)\,,\\
\frac{\Delta}{\rho}& = &1 + O(x)\,
\end{eqnarray} 
\noindent
where $x=\rho/R_g^2$ and $\Phi_0$ is the (generally non-zero) value of
the dilaton field at the origin. Recall that we have fixed an overall
scale by imposing the condition $\Phi(\infty)=0$, after what the length
parameter $R_g$ in the Eqs.(21)-(23) have been absorbed by passing to a
dimensional variable $x$. The system consists of two equations of the
second order and one of the first order, hence the solution will be fixed
completely by the boundary conditions for $f(0)$, $f'(0)$, $\Phi(0)$,
$\Phi'(0)$ and $\Delta(0)$, which are parameterized according to Eqs.
(25)-(27) in terms of $b$ and $\Phi_0$. Alternatively, the problem can be
thought of as the Stourm-Liouville problem with fixed $f(0)=-1$,
$\Delta/\rho \rightarrow 1$ as $\rho \rightarrow 0$, $|f(\infty)|=1$,
$\Phi(\infty)=0$ and $\Delta/\rho \rightarrow 1$ as $\rho
\rightarrow \infty$. The solution, like Bartnik-McKinnon solution
\cite{r4} for the pure EYM system, exists for discrete values of the
parameters $b$ and $\Phi_0$, labeled by the number of zeros $n$ of
the Yang-Mills function $f$. For each $n$ there exist one pair of
values of $b$ and $\Phi_0$, hence the family of solutions remains
one-parametric as in the EYM case \cite{r4}. These values found
numerically for some lower $n$ are shown in the tab.1 together with
the corresponding values of the total mass and the dilaton charge
as given by Eqs. (18), (19). With increasing $n$ both these quantities
are likely to tend to some limiting values. Numerical solutions for
$f$, $\Phi$, and $\sigma$ are shown at the figs.1-3,
behavior of the mass function is depicted at the fig.4. Note that
functions $\Phi(r)$ and $\sigma(r)$ are monotonic as can be
anticipated from the Eqs.(20)-(23). It is worth to note that like
in the abelian case \cite{r1} a dilaton charge is not an independent
parameter. Note that numerically $M$ and $D$ (in the units $R_g=1$)
are very close together.
\begin{table}[htb]\begin{center}
\centerline{\small\bf Table 1}
\medskip

\begin{tabular}{|c|c|c|c|c|c|}       \hline
$n$ & $b$ & $\Phi _0$ & $\sigma _0$ & $M$ & $D$ \\  \hline
1 & 1.0718 & 0.9300 & 0.3936 & 0.5777 & 0.5782 \\
2 & 8.3612 & 1.7923 & 0.1665 & 0.6850 & 0.6852 \\
3 & 53.8351 & 2.6320 & 0.0678 & 0.7035 & 0.7042 \\ \hline
\end{tabular}\end{center}\end{table}

Physical interpretation of the regular solutions can be given
along the lines of \cite{r8}. A path in the functional space
connecting the Yang-Mills vacua with neighbouring winding numbers
can be constructed using the parameterization of the Yang-Mills
connection similar to given in \cite{r8} adding to it
an appropriate dilaton path. Odd-$n$ solutions then
can be shown to play the role of sphalerons. For all $n$ the solutions
are expected to be unstable with one of the instability modes being
the rolling down mode from the top of the potential barrier which
separates the Yang-Mills vacua.

In the black hole case an additional continuous real parameter
$\rho_H = r_H^2$, with $r_H$ being the largest (simple) zero of the
function $\Delta$, introduces an independent length scale. The whole
family of solutions will then be two-parametric in terms of $(r_H, n)$,
$x=\rho/\rho_H$ will be a new radial variable. A series expansion of
the solution in the outside vicinity of the horizon reads
\begin{eqnarray}
f&=&f_H + \frac{f_H(f_H^2-1)}{2G_H}\;y \,+\,O(y^2)\,,\\
\Phi&=&\Phi_H + \frac{y}{2}(1-\frac{1}{G_H})  + O(y^2)\,,\\
\Delta&=& \frac{y}{2}\, \rho_H G_H\;(1+O(y))\,,
\end{eqnarray} 
\noindent
where $G_H=G(\rho_H)$ and $y=x-1$. The solutions have different
character depending on the value of the parameter $r_H$.
Likely to the pure EYM case \cite{r5}, in the limit $r_H\rightarrow 0$
the solution outside the horizon is indistinguishable from the regular
one. In the opposite limit $r_H\rightarrow \infty$ the Yang-Mills
equation decouples from the coupled gravity-dilaton system, and likely
to the pure EYM case one could anticipate the existence of the (fully
non-linear) Yang-Mills equation on the black hole background. It can
be easily seen from the expansion (29) that the derivative of the
dilaton field at the horizon tends to zero and hence in this limit
the dilaton in not excited at all (otherwise regular black hole
solution with dilaton but without Yang-Mills hair could
not exist). So in the limit $r_H \rightarrow \infty$ the solution
reduces to that of the EYM system \cite{r6} (in particular, an
analytic form of the solution for $n=1$ is known).

For $r_H \sim 1$ the solution was studied numerically. With fixed
$x_H$ it exists for the discrete values of the parameters $f_H$
and $\Phi_H$. These values for $r_H=1$ are shown in the tab. 2
(with the same meaning of $n$ as before) together with the horizon
values of $\sigma$,  the field mass $M-M_H$ and the dilaton charge.
Figures 5--7 depict the corresponding numerical curves. The similarity
with the EYM black holes  \cite{r5} is manifest. Both the field mass
and the dilaton charge are rapidly saturated with increasing $n$.
\begin{table}[htb]\begin{center}
\centerline{\small\bf Table 2}
\medskip

\begin{tabular}{|c|c|c|c|c|c|c|c|}\hline
$n$ & $f_H$ & $\Phi _H$ & $\sigma _H$ & $M$ & $M-M_H$ & $D$ & $T/T_H$ \\ \hline
1 & -0.5937 & 0.4420 & 0.7850 & 0.8368 & 0.3368 & 0.5124 & 0.6490\\
2 & -0.1321 & 0.5445 & 0.8598 & 0.8651 & 0.3651 & 0.5749 & 0.5804\\
3 & -0.0218 & 0.5493 & 0.8653 & 0.8658 & 0.3658 & 0.5765 & 0.5771\\ \hline
\end{tabular}\end{center}\end{table}

To illustrate the difference between the metrics of the non-abelian
and the abelian magnetically charged solution \cite{r1} one can define
the metric ''magnetic charge`` function $P^2(r)$ as follows
\begin{equation}
g_{00}=1-{2M/r}\sqrt{1+D^2/r^2} +P^2/r^2\, ,
\end{equation} 
\noindent
such that $P={\rm const}$ (magnetic charge) in the abelian case. The
corresponding curves for non-abelian stringy black holes are shown
at the fig.~8.

We have also calculated the Hawking temperature
\begin{eqnarray}
T&=& \frac{\sigma (r_H)\,(\frac{d\Delta}{dr})_{r=r_H}}{4\pi r_H^2}
\nonumber\\ &=& \frac{\sigma (r_H)\,G_H}{4\pi r_H}\,.
\end{eqnarray} 
\noindent
The ratio of the Hawking temperature to the temperature of the
Schwarzschild black hole posessing the same radius of the event
horizon $T_H=1/4\pi r_H$ is shown in the tab. 2. We will discuss
the thermodynamics of non-abelian stringy black holes in a
separate publication.

To summarize: we have found strong numerical evidence in favour of
the existence of both regular and black hole 4-dimensional
solutions of the heterotic string effective action in the SU(2)
sector. They generalize the corresponding EYM sphalerons and
non-abelian black holes, are unstable, and are supposed to play
a similar role in the context of the quantum theory. They have
quantized values of the dilaton charge and interact among
themselves by both gravitational and dilaton long-range forces.
Spherically symmetric solutions have zero both Lorentz and Yang-Mills
Chern-Simons 3-forms and, correspondingly, a Kalb-Ramond field
is not excited (though topological axion hair is not excluded).


\newpage
{\Large Figure Captions}

Fig. 1. Yang-Mills magnetic field function $f$, regular case.

Fig. 2. Dilaton field, regular case.

Fig. 3. Metric function $\sigma$, regular case.

Fig. 4. Mass distribution for $n=2$ regular solution: A -- total mass,
        B -- Yang-Mills contribution (second term in the Eq. (19)),
        C -- dilaton contribution (first term in the Eq. (19)).

Fig. 5. Yang-Mills function $f$ for the $r_H=1$ black holes.

Fig. 6. Dilaton field for $r_H=1$ black hole.

Fig. 7. Metric function $\sigma$ for $r_H=1$ black hole.

Fig. 8. Metric ''charge`` function $P^2(r)$.

\end{document}